\documentclass[aps,
               prl,
               reprint,
               groupedaddress, 
               showpacs,
               citeautoscript]{revtex4-1}

\usepackage{amsmath}
\usepackage{amssymb}
\usepackage{graphicx}

\newcommand{\op}[1]{\hat{#1}}

\newcommand{\up}{\uparrow}
\newcommand{\down}{\downarrow}

\begin{document}

\title{Optimizing the Cooper pair splitting efficiency in
       a Y-shaped junction}

\author{K.J. Pototzky and E.K.U. Gross} 

\affiliation{Max Planck Institute of Microstructure Physics, 
             06120 Halle (Saale), Germany}

\date{\today}

\begin{abstract}
    This letter is devoted to the
    optimization of the Cooper pair splitting efficiency
    in a Y-shaped junction. The latter
    consists of 
    two quantum dots,
    one superconducting and two normal leads.
    We tailor the bias in the two normal leads
    such that the Cooper pairs leaving
    the superconductor are split up
    resulting in entangled electrons,
    one on each quantum dot. 
    We are able to achieve a 
    splitting efficiency of more than 99\%
    which is significantly better than
    the efficiencies
    obtained in experiments so far.
\end{abstract}

\pacs{
      74.45.+c 
      03.67.Bg 
      73.63.-b  
      73.63.Kv  
     }

\maketitle

\begin{figure}[htb]
            \begin{center}
                \includegraphics{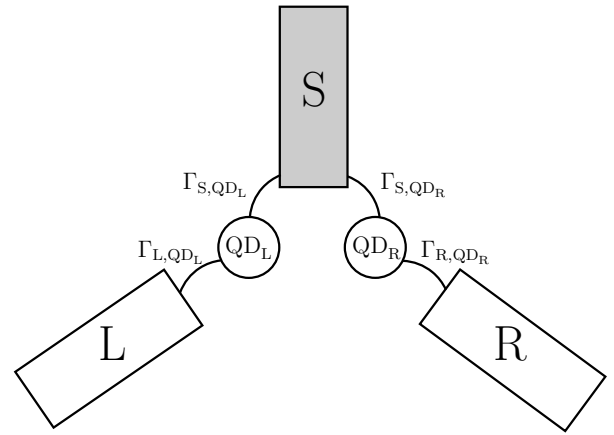}
                \caption{Sketch of the Y junction 
                         consisting of one superconducting ($\textnormal{S}$)
                         and two normal-conducting leads ($\textnormal{L}$ and $\textnormal{R}$)
                         as well as two quantum dots 
                         ($\textnormal{QD}_{\textnormal{L}}$ and $\textnormal{QD}_{\textnormal{R}}$).
                         The coupling strengths 
                         $\Gamma_{\alpha, \textnormal{QD}_\beta}$
                         associated with the various links are 
                         indicated as well.
                         }
                \label{fig:Sketch-Y-junction}
            \end{center}
        \end{figure}

    The entanglement of
    quantum particles has 
    fascinated the scientific community
    since the proposition of the Einstein-Podolsky-Rosen
    Gedankenexperiment \cite{Einstein1935}. It is
    directly linked to
    the question of
    non-locality of 
    quantum mechanics. 
    A violation of Bell's inequality
    would prove the latter \cite{Bell1964}. 
    Great progress has been 
    achieved with entangled photons,
    but the final experiment 
    ruling out all possible loopholes
    has not yet been accomplished \cite{Giustina2013}.
    To do similar experiments
    with electrons
    is much more difficult and 
    remains an open challenge.
    In recent years,
    a number of ingenious 
    experiments to create
    entangled
    electrons have been performed
    \cite{Hofstetter2009,Herrmann2010,Schindele2012},
    going along with several 
    theoretical developments \cite{Recher2001,Recher2002,Sauret2004,Morten2006,Golubev2007}.
    The basic idea is to use 
    a superconductor as a
    source of entangled electrons.
    In the BCS ground state, 
    electrons form Cooper pairs
    due to the attractive interaction
    caused by phonons. These pairs
    consist of two electrons 
    with opposite spin and momentum.
    The idea is to split the Cooper pairs
    by making them leave
    the superconducting lead, forcing
    one electron to move to a 
    quantum dot on the left    
    and the other to a quantum 
    dot on the right (see sketch in Fig. \ref{fig:Sketch-Y-junction}).
    From these two quantum dots,
    the electrons are transported 
    further into two metallic 
    (normal-conducting) leads,
    $\textnormal{L}$ and $\textnormal{R}$,
    where they get spatially separated.
    Since the splitting process does not 
    affect the spin, the electrons
    are entangled because they stay
    in a spin-singlet state while
    separating.
    However, this process
    competes with the case
    of both electrons moving into the same lead.
    The latter can be suppressed 
    by a large charging energy 
    of the quantum dots caused by
    the Coulomb interaction. 
    This make double occupancies less likely.
    The splitting further benefits from a
    weak coupling of the
    quantum dots to the leads
    compared to the superconducting gap \cite{Hofstetter2009}.
    Splitting efficiencies up to
    $90\%$ have been 
    realized in recent experiments \cite{Schindele2012}
    being significantly higher than previous results.
    Despite this progress,
    the experimental proof 
    of the violation of
    Bell's inequality 
    is still pending. 
    
    In this letter, we 
    propose a way 
    to achieve splitting 
    efficiencies of $99\%$ and more,
    which we hope will 
    help the eventual experimental
    demonstration of the violation
    of Bell's
    inequality. 
    The traditional approach to achieve 
    high splitting rates relies
    on a large Coulomb repulsion
    on the quantum dots. The 
    approach proposed in this letter
    is different: Our strategy
    is to use optimal control theory
    to tailor the bias in the normal 
    leads in such a way that the splitting
    probability is maximized.
            
        In order to describe transport
        processes through the
        Y-shaped junction sketched in Fig. \ref{fig:Sketch-Y-junction}
        we consider the
        following model Hamiltonian:
        \begin{align}
            \label{eqn:sec2:full-Hamiltonian}
            \op{H}(t)            &=  \sum_{\alpha \in \{\textnormal{L},\textnormal{R},\textnormal{S}\}} \op{H}_{\alpha} + \sum_{\alpha \in \{\textnormal{L},\textnormal{R},\textnormal{S}\}}\op{H}_{T,\alpha}(t),\\
            \op{H}_{\alpha}      &= \sum_{k=0}^\infty \sum_{\sigma\in\{\up,\down\}}\left(t_\alpha \op{c}_{\alpha k\sigma}^\dagger \op{c}_{\alpha(k+1)\sigma} + \textnormal{h.c.} \right)\\
                                 &  \nonumber + \quad \sum_{k=0}^\infty \left(\Delta_\alpha \op{c}_{\alpha k\up}^\dagger \op{c}_{\alpha k\down}^\dagger + \textnormal{h.c.} \right) \textnormal{ for }\alpha \in \{ \textnormal{S}, \textnormal{L}, \textnormal{R} \}, \nonumber \\
            \op{H}_{T,S}         &= \sum_{\alpha \in \{\textnormal{L},\textnormal{R}\}}\sum_{\sigma \in \{\up,\down\}}\left( t_{\textnormal{S}, \textnormal{QD}_\alpha} \op{c}_{\textnormal{S}0\sigma}^\dagger \op{d}_{\textnormal{QD}_\alpha \sigma} + \textnormal{h.c.} \right),
        \end{align}    
        \begin{align}
            \op{H}_{T,\alpha}(t) &= \sum_{\sigma \in \{\up,\down\}}\left( t_{\alpha, \textnormal{QD}_\alpha} e^{i\gamma_{\alpha, \textnormal{QD}_\alpha}(t)}\op{c}_{\alpha0\sigma}^\dagger \op{d}_{\textnormal{QD}_\alpha \sigma} + \textnormal{h.c.} \right) \nonumber \\
                                 & \qquad \qquad \qquad \textnormal{for }\alpha \in \{ \textnormal{L}, \textnormal{R} \}
            \label{eqn:sec2:last-eqn-full-Hamiltonian}
        \end{align}
        with the Peierls' phases $\gamma_{\alpha, \textnormal{QD}_\alpha}(t) = \int_0^t\,dt' U_\alpha(t')$ and the biases
        $U_\alpha(t), \alpha \in \{\textnormal{L},\textnormal{R}\}$.
        The operator $\op{c}_{\alpha k\sigma}^\dagger$ ($\op{c}_{\alpha k\sigma}$) creates (annihilates) 
        an electron at site $k \in \mathbb{N}$ in the lead $\alpha \in \{ \textnormal{S}, \textnormal{L}, \textnormal{R}\}$ with spin $\sigma \in \{\up,\down\}$.
        The operator $\op{d}_{\textnormal{QD}_\alpha \sigma}^\dagger$ ($\op{d}_{\textnormal{QD}_\alpha \sigma}$) represents the creation (annihilation) 
        of an electron on the quantum dot $\alpha \in \{\textnormal{L}, \textnormal{R}\}$.
        
        All parameters in equations (\ref{eqn:sec2:full-Hamiltonian}) - (\ref{eqn:sec2:last-eqn-full-Hamiltonian}) are chosen real and positive.
        We shall work at temperature $T=0$ and assume the wide band limit $t_{\alpha, \textnormal{QD}_\beta} \ll t_\alpha $.
        In this limit, the results only depend on 
        the ratios
        $\Gamma_{\alpha, \textnormal{QD}_\beta} = {2t_{\alpha, \textnormal{QD}_\beta}^2}/{t_\alpha}$
        but not on the hopping elements individually. 

        The pairing potentials can be written as $\Delta_\alpha = \xi_\alpha \widetilde{\Delta}$
        which allows a dimensionless representation of the problem by measuring
        times in units of $\widetilde{\Delta}^{-1}$ and 
        energies in units of $\widetilde{\Delta}$. 
        We set $\xi_\textnormal{S}=1$
        for the superconducting lead $\textnormal{S}$ 
        and $\xi_\textnormal{L}=\xi_\textnormal{R}=0$
        for the other two.
        Due to the presence of superconductivity, we have to solve the
        time-dependent Bogoliubov-de Gennes equation, which is 
        a Schr\"odinger-like equation in electron-hole space.
        For the single particle wave functions
        $\psi_q(k,t) = [u_q(k,t), v_q(k,t)]^t$ 
        it reads as follows:
        \begin{align}
            \label{eqn:BdG-equation}
            i\frac{\,d}{\,dt}
            \left(
                \begin{matrix}
                    u_q(k,t)\\
                    v_q(k,t)
                \end{matrix}
            \right)
                &=
            \sum_l
            \mathbf{H}_{kl}(t)
            \left(
                \begin{matrix}
                    u_q(l,t)\\
                    v_q(l,t)
                \end{matrix}
            \right),\\
            \mathbf{H}_{kl}(t) &=
            \left(
                \begin{matrix}
                    \mathbf{h}_{kl}(t)           & \mathbf{\Delta}_{kl} \\
                    \mathbf{\Delta}_{kl}^\dagger & -\mathbf{h}_{kl}^\dagger(t)
                \end{matrix}
            \right).
        \end{align}
        The algorithm for the time propagation of the single particle
        wave functions $\psi_q(k,t)$ as well as the 
        initial state calculation is explained in the
        work of Stefanucci \textit{et. al.} \cite{Stefanucci2010},
        which extends the method of Kurth \textit{et. al.} \cite{Kurth2005}
        to superconducting leads. The initial
        state is chosen to be the ground state
        of the system.

\begin{figure}[htb]
            \begin{center}
                \vspace*{0.5cm}
                \includegraphics{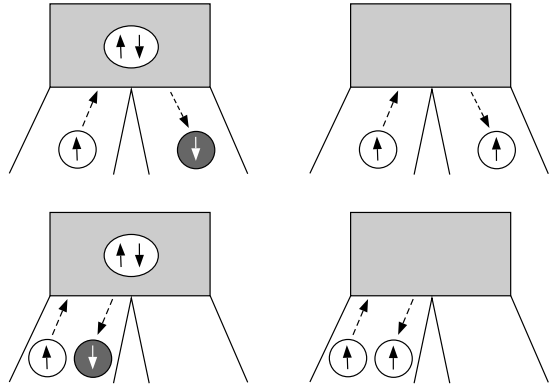}
                \caption{Overview of the four possible reflection processes.
                         Black arrows indicate electrons, white arrows
                         represent holes. The gray block is the superconducting lead 
                         $\textnormal{S}$ of Fig. \ref{fig:Sketch-Y-junction}.
                         Top left: Sketch of a crossed Andreev reflection.
                         The incoming spin up electron in the left lead gets 
                         reflected as a spin down hole to the right lead.
                         Simultaneously, a Cooper pair is created in the 
                         superconducting lead. The opposite process, which
                         removes a Cooper pair from the superconductor,
                         is also possible.
                         Bottom left:
                         The reflected hole stays in the left lead.
                         This corresponds to the normal Andreev reflection.
                         Top right: Sketch of an elastic cotunneling process.
                         Now, the incoming electron gets reflected into 
                         the right lead. 
                         Bottom right: Alternatively, the 
                         electron can also be reflected into the 
                         left lead corresponding to normal reflection.}
                \label{fig:Crossed-Andreev-reflection}
            \end{center}
        \end{figure}

        In the following, we demonstrate how to
        optimize the Cooper pair splitting
        efficiency in the above model
        of a two-quantum dot Y-junction. 
        The goal is to 
        operate the device as
        a Cooper pair splitter
        that creates
        entangled electrons
        on the two quantum dots.
        The splitting of a Cooper pair 
        can be understood
        as a crossed Andreev reflection.
        An incoming electron in one of the
        normal leads gets reflected
        into the other lead as a hole.
        This creates a Cooper pair 
        in the superconductor. The
        process is sketched in Fig. \ref{fig:Crossed-Andreev-reflection} (top left).
        Similarly, the opposite process 
        removes a Cooper pair from the
        superconductor. 
        Besides, there 
        are three other possible
        reflection processes: (a) normal reflection,
        (b) Andreev reflection,
        and (c)
        elastic cotunneling.
        The latter corresponds to
        a reflection of the incoming electron
        to the opposite lead.
        These three processes together with 
        the crossed Andreev reflection 
        are all sketched in
        Fig. \ref{fig:Crossed-Andreev-reflection}. 
        
        The central ingredient for the 
        optimization
        process is the proper definition
        of a suitable 
        objective function
        which is then to be maximized.
        It has 
        to quantify the Cooper pair splitting
        efficiency. 
        To this end, we first define
        the so-called
        pairing density or anomalous density as
        \begin{align}
            P_{\textnormal{QD}_\alpha,\textnormal{QD}_\beta}(t) &= \langle \hat{d}_{\textnormal{QD}_\alpha\downarrow,\textnormal{H}}(t)\hat{d}_{\textnormal{QD}_\beta\uparrow,\textnormal{H}}(t)\rangle.
        \end{align}
        We use its absolute value
        squared $|P_{\textnormal{QD}_\alpha,\textnormal{QD}_\beta}(t)|^2$ 
        as a measure 
        for the Cooper pair density
        with one electron at $\textnormal{QD}_\alpha$
        and the other at $\textnormal{QD}_\beta$.
        We propose to maximize 
        the following 
        objective function:
        \begin{widetext}
            \begin{align}
                \label{eqn:objective-function}
                \frac{1}{t_1-t_0}\int_{t_0}^{t_1} \,dt \frac{|P_{\textnormal{QD}_\textnormal{L},\textnormal{QD}_\textnormal{R}}(t)|  ^2+|P_{\textnormal{QD}_\textnormal{R},\textnormal{QD}_\textnormal{L}}(t)|^2}
                    {|P_{\textnormal{QD}_\textnormal{L},\textnormal{QD}_\textnormal{L}}(t)|^2+|P_{\textnormal{QD}_\textnormal{L},\textnormal{QD}_\textnormal{R}}(t)|^2+|P_{\textnormal{QD}_\textnormal{R},\textnormal{QD}_\textnormal{L}}(t)|^2+|P_{\textnormal{QD}_\textnormal{R},\textnormal{QD}_\textnormal{R}}(t)|^2}.
            \end{align}
        \end{widetext}
        The fraction represents the
        Cooper pair splitting efficiency at time t, which is expressed 
        as the amount of Cooper pairs being split up divided 
        by the total amount of Cooper pairs on the quantum dots. 
        We calculate its average over the time span 
        from $t_0$ to $t_1$.
        The pairing densities $P_{\textnormal{QD}_\alpha,\textnormal{QD}_\beta}(t)$
        are obtained from the single particle wave functions $\psi_q(t)$, i.e.,
        the solutions of the time-dependent 
        Bogoliubov-de Gennes equation (\ref{eqn:BdG-equation}).

        We want to tailor the time-dependent bias such that
        the time averaged 
        Cooper pair splitting
        efficiency, i.e. the objective function (\ref{eqn:objective-function}), is maximized.
        The numerical set-up to do this is by representing
        $U_\alpha(t)$ by cubic 
        splines with $N+1$ equidistant nodes 
        at $\tau_k = \frac{k}{N}T, k\in \{0,\ldots, N\}$.
        We choose $\frac{\,d}{\,dt}U_\alpha(\tau_0) = \frac{\,d}{\,dt}U_\alpha(\tau_N)=0$
        as boundary conditions for the splines.
        The dependence of the Hamiltonian on the bias $U_\alpha(t)$ is replaced by
        a dependence on the vector
        \begin{equation}
            U_\alpha(t) \to \left[U_\alpha(\tau_0), \ldots, U_\alpha (\tau_{N})\right] \equiv \vec{u}_\alpha.
        \end{equation}

        The bias $U_\alpha(t)$ becomes a function of $\vec{u}_\alpha$,
        namely $U_\alpha(\vec{u}_\alpha, t)$.        
        This then yields a standard non-linear optimization problem 
        with unknown variables $U_\alpha(\tau_k)$. We further impose
        the condition $U_\alpha(\tau_0)=0$ since the 
        bias has to be continuous and we assume $U_{\alpha}(t<0)=0$. 

        The corresponding optimization problem then reads
        \begin{widetext}
            \begin{align}
                &\max_{\vec{u}_{\textnormal{L}}, \vec{u}_{\textnormal{R}} \in \mathbb{R}^{N+1}} 
                    \frac{1}{t_1-t_0}\int_{t_0}^{t_1}\,dt \frac{|P_{\textnormal{QD}_\textnormal{L},\textnormal{QD}_\textnormal{R}}(t)|^2+|P_{\textnormal{QD}_\textnormal{R},\textnormal{QD}_\textnormal{L}}(t)|^2}
                    {|P_{\textnormal{QD}_\textnormal{L},\textnormal{QD}_\textnormal{L}}(t)|^2+|P_{\textnormal{QD}_\textnormal{L},\textnormal{QD}_\textnormal{R}}(t)|^2+|P_{\textnormal{QD}_\textnormal{R},\textnormal{QD}_\textnormal{L}}(t)|^2+|P_{\textnormal{QD}_\textnormal{R},\textnormal{QD}_\textnormal{R}}(t)|^2}\\
                &\begin{array}{rcl}
                    \textnormal{with } \quad P_{\textnormal{QD}_\alpha, \textnormal{QD}_\beta} &=& \int \,dq f(\epsilon_q) u_q(\textnormal{QD}_\alpha, t) v_q(\textnormal{QD}_\beta, t)^\star,\\
                    \quad i\partial_t \psi_q(t)   &=& \mathbf{H} (\vec{u}_{\textnormal{L}}, \vec{u}_{\textnormal{R}}, t)\psi_q(t), \quad t \in [0, T],\\ 
                    \psi_q(0)                                        &=& \psi_q^0,\\     
                    U_\alpha(\vec{u}_\alpha, \tau_0)                 &=& 0,\quad  \alpha \in \{\textnormal{L}, \textnormal{R}\}.
                \end{array} \nonumber
            \end{align}
        \end{widetext}

        The problem can be solved using standard 
        derivative-free
        algorithms for non-linear optimization problems.
        This approach has already been used in 
        several other works \cite{Castro2009, Krieger2011, Hellgren2013, Raesaenen2013}.
        We use the algorithm BOBYQA \cite{BOBYQA} 
        provided by the library NLopt \cite{NLopt_paper}. 
        It outperforms all other tested 
        optimization algorithms.

        To achieve high splitting efficiencies it 
        is essential that the
        junction is asymmetric,
        i.e. the couplings to the left 
        and to the right quantum dot must not be equal.
        This is necessary since we observe
        an upper bound of $50\%$ for the Cooper
        pair splitting efficiency
        in symmetric junctions,
        which is already achieved in 
        the ground state
        by the usual Cooper pair 
        tunneling leading to the proximity effect. Hence any
        optimization starting in the ground
        state will not improve
        the results. We therefore
        choose an asymmetric coupling
        of the quantum dots to the normal leads.
        
        \begin{figure}[htb]
            \begin{center}
                \includegraphics[]{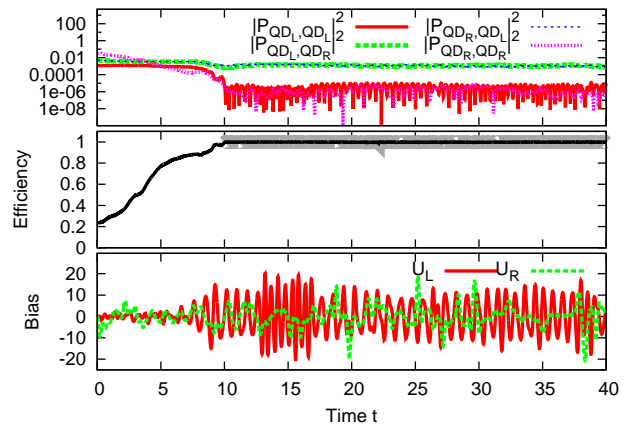}
                \caption{Simulation with an optimized bias. 
                         Upper panel: $|P_{\textnormal{QD}_\alpha, \textnormal{QD}_\beta}(t)|^2$
                         as a function of time. 
                         Middle panel: Resulting efficiency,
                         gray line indicates time interval of 
                         optimization.
                         Lower panel: Tailored bias $U_\textnormal{L}(t)$
                         and $U_\textnormal{R}(t)$ of the optimization.
                         The parameters are: 
                         $\Gamma_{\textnormal{S},\textnormal{QD}_\textnormal{L}} = \Gamma_{\textnormal{S},\textnormal{QD}_\textnormal{R}} = \Gamma_{\textnormal{N},\textnormal{QD}_\textnormal{L}}=0.2$,
                         $\Gamma_{\textnormal{N},\textnormal{QD}_\textnormal{R}}=1$, $N=200$.}
                \label{fig:efficiency-optimized+constraint}
            \end{center}
        \end{figure}
        
        The results of such an optimization are depicted in
        Fig. \ref{fig:efficiency-optimized+constraint}.
        The bias is tailored such that the Cooper pair 
        splitting efficiency is maximized. It suppresses the non-splitting 
        processes. The efficiency
        is optimized in the time interval
        from $t_0=10$ to $t_1=40$.
        This interval is indicated by the underlying
        thick gray line in the
        plot of the efficiency (middle).
        In this interval, we 
        achieve an average efficiency of 
        more than $99\%$.
        The values of $|P_{\textnormal{QD}_\textnormal{L},\textnormal{QD}_\textnormal{R}}(t)|^2$
        and  $|P_{\textnormal{QD}_\textnormal{R},\textnormal{QD}_\textnormal{L}}(t)|^2$
        are on top of each other.
        This result demonstrates that the
        Coulomb interaction at the quantum dots
        is not necessary in order to 
        obtain high efficiencies. One
        can also succeed with optimized
        biases.

        To summarize, we have demonstrated how to optimize
        the Cooper pair splitting 
        efficiency in a Y-shaped junction
        by suitably tailoring the bias.
        In this way, we are able to achieve splitting
        efficiencies of
        $99\%$ and more, which is 
        significantly higher than present
        experiments.
        This efficiency may help to finally demonstrate
        a violation of Bell's inequality with electrons.


\bibliography{BibtexDatabase}

\end{document}